\documentclass[conference]{IEEEtran}
\IEEEoverridecommandlockouts
\usepackage{amsmath, amssymb, amsfonts}
\usepackage{graphicx}
\usepackage[linesnumbered,ruled,vlined]{algorithm2e}
\usepackage{tikz}
\usepackage{amsmath}
\usepackage{booktabs}
\usetikzlibrary{positioning, shapes.geometric}
\usepackage{pgfplots}
\pgfplotsset{compat=1.18} % or latest you have installed
\definecolor{darkblue}{RGB}{0,0,139}
\definecolor{darkred}{RGB}{139,0,0}
\SetKwInput{KwInput}{Input}
\SetKwInput{KwOutput}{Output}
\usepackage{algorithmic}
\usepackage{cite}
\usepackage{bbm}

\usepackage{booktabs} % For professional tables
\usepackage{hyperref} % For hyperlinks
\usepackage{cleveref}
\usepackage{xcolor}
\usepackage{enumitem}
\usepackage{tikz}
\usepackage{adjustbox}
\usepackage{eqparbox}
\usepackage{listings}
\usepackage[caption=false,font=footnotesize]{subfig}

\usepackage{float}
\usepackage{enumitem}
\usepackage{multirow}
\usepackage{amsmath}
\usepackage{amsthm}
\usepackage{tabularx}
\usepackage[acronym]{glossaries}

\definecolor{greensquare}{rgb}{0.0, 0.6, 0.0}
\definecolor{DarkMaroon}{RGB}{128, 0, 0}
\definecolor{Mustard}{RGB}{255, 204, 0}
\definecolor{Blue}{RGB}{0, 0, 255}

\newacronym{6g}{6G}{Sixth Generation}
\newacronym{uav}{UAV}{Unmanned Aerial Vehicle}
\newacronym{isac}{ISAC}{Integrated Sensing and Communication}
\newacronym{ris}{RIS}{Reconfigurable Intelligent Surface}
\newacronym{ntn}{NTN}{Non-Terrestrial Network}
\newacronym{jcas}{JCAS}{Joint Communication and Sensing}
\newacronym{pls}{PLS}{Physical Layer Security}
\newacronym{hndl}{HNDL}{Harvest-Now, Decrypt-Later}
\newacronym{pqc}{PQC}{Post-Quantum Cryptography}
\newacronym{glrt}{GLRT}{Generalised Likelihood Ratio Test}
\newacronym{nist}{NIST}{National Institute of Standards and Technology}
\newacronym{3gpp}{3GPP}{3rd Generation Partnership Project}
\newacronym{siu}{SIU}{Secure ISAC Utility}
\newacronym{qrtm}{QRTM}{Quantum-Resilient Threat Modelling}
\newacronym{gNB}{gNB}{Next-Generation Node B}
\newacronym{mlkem}{ML-KEM}{Module Lattice–Key Encapsulation Mechanism}
\newacronym{falcon}{Falcon}{Fast-Fourier Lattice-Based Compact Signatures over NTRU}
\newacronym{shor}{Shor}{Shor’s Quantum Factoring Algorithm}
\newacronym{grover}{Grover}{Grover’s Quantum Search Algorithm}
\newacronym{atis}{ATIS}{Alliance for Telecommunications Industry Solutions}
\newacronym{5ga}{5GA}{5G Americas}
\newacronym{rf}{RF}{Radio Frequency}
\newacronym{ecc}{ECC}{Elliptic Curve Cryptography}
\newacronym{qgan}{QGAN}{Quantum Generative Adversarial Network}
\newacronym{rsa}{RSA}{Rivest–Shamir–Adleman}
\newacronym{c2}{C2}{Command and Control}
\newacronym{sa}{SA}{Situational Awareness}
\newacronym{leo}{LEO}{Low Earth Orbit}
\newacronym{qkd}{QKD}{Quantum Key Distribution}

\begin{document}
\title{Quantum-Resilient Threat Modelling for Secure RIS-Assisted ISAC in 6G UAV Corridors
\thanks{The authors gratefully acknowledge Liverpool John Moores University in the United Kingdom, the Interdisciplinary Research Centre for Aviation and Space Exploration (IRC-ASE) at KFUPM, Saudi Arabia, and the Department of Applied Sciences at Universiti Teknologi PETRONAS, Malaysia, for their collaborative support of this research project.}}

\author{
\IEEEauthorblockN{Sana Hafeez}
\IEEEauthorblockA{\textit{Digital Technologies \& Artificial Intelligence}\\
Digital Innovation Research Institute,\\
Liverpool John Moores University, UK\\
S.Hafeez@ljmu.ac.uk}
\and
\IEEEauthorblockN{Ghulam E Mustafa Abro}
\IEEEauthorblockA{\textit{Interdisciplinary Research Centre for}\\
 Aviation \& Space Exploration (IRC-ASE),\\
 KFUPM Dhahran, 31261, Saudi Arabia\\
Mustafa.abro@ieee.org}
\and
\IEEEauthorblockN{Hifza Mustafa}
\IEEEauthorblockA{\textit{Department of Applied Sciences}\\
Universiti Teknologi PETRONAS,\\
Seri Iskandar, 32610, Perak Malaysia\\
Mustafahifza@gmail.com}
}
\maketitle
\IEEEpubid{\makebox[\columnwidth]{979-8-3315-6576-3/25/\$31.00~\copyright~2025 IEEE \hfill}
\hspace{\columnsep}\makebox[\columnwidth]{}}
\begin{abstract}
The swift implementation of unmanned aerial vehicle (UAV) corridors in sixth-generation (6G) networks necessitates safe, intelligence-driven integrated sensing and communications (ISAC). Reconfigurable intelligent surfaces (RIS) improve spectrum efficiency, localisation precision, and situational awareness, while also introducing new vulnerabilities. The emergence of quantum computing heightens hazards associated with harvest-now, decrypt-later tactics and quantum-enhanced spoofing. We propose a \textit{quantum-resilient threat modelling (QRTM)} framework for RIS-assisted ISAC in UAV corridors to tackle these problems. QRTM integrates classical, quantum-ready, and quantum-aided adversaries, addressing them with post-quantum cryptographic (PQC) primitives: ML-KEM for key establishment and Falcon for authentication, both incorporated inside RIS control signalling and UAV coordination. To enhance security sensing, we present RIS-coded scene watermarking validated by a generalised likelihood ratio test (GLRT), with its detection probability characterised by a Marcum-$Q$ function. Additionally, we establish a secure ISAC utility (SIU) that concurrently optimises secrecy rate, spoofing detection, and throughput within RIS limitations, facilitated by a scheduler with $\mathcal{O}(n^2)$ complexity. Monte Carlo evaluations utilising 3GPP Release-19 mid-band urban-canyon models (7–15 GHz) reveal spoof-detection probability approaching 0.99 at $P_{\mathrm{FA}}=10^{-3}$, secrecy-rate retention surpassing 90\% versus quantum-capable adversaries, and signal interference utilisation enhancements of around 25\% relative to baselines. These findings underscore a standards-compliant approach to establishing a reliable, quantum-resilient ISAC for UAV corridors in smart cities and non-terrestrial networks.
\end{abstract}
\IEEEpubidadjcol
\section{Introduction}
\subsection{Background}
\IEEEPARstart{Q}{uantum} dynamics are transforming the security underpinnings of wireless systems as sixth generation amalgamates communication, sensing, and intelligence into cohesive air-ground infrastructures. A notable instance is the establishment of \gls{uav} corridors designated as airborne routes facilitating swift logistics, medical, and emergency services. In contrast to terrestrial \gls{isac}, these corridors necessitate continuous line-of-sight, elevated mobility, and decentralised coordination, rendering them susceptible to eavesdropping, spoofing, and interference. Adversaries can readily exploit unsecured aerial connections, jeopardising both control and sensing channels. Augmented by \gls{ris}, \gls{uav} corridors provide sub-meter localisation, optimised spectrum utilisation, and enhanced situational awareness \cite{ref1}. Standardisation embodies this: 3GPP Release 19 presents midband (7–24 GHz) UAV/NTN models \cite{ref2}, whereas ATIS Phase II revises TR 38.901 \cite{ref3, ref4}. Similarly, 5GA emphasises \gls{ris}, \gls{jcas}-driven spectrum sharing as essential enablers of 6G.

\subsection{Research Limitations and Gaps}
Important restrictions still exist despite advancements. Most works on \gls{isac} and \gls{pls} continue to rely on traditional assumptions \cite{ref5,ref6} utilise \gls{ris} to enhance secrecy or mitigate eavesdropping, while \cite{ref7} investigates \gls{ris}-based \gls{rf} fingerprinting; however, none address quantum adversaries. Quantum computing transforms threats: Grover’s quadratic speedup \cite{ref8} undermines symmetric search, Shor’s factoring \cite{ref9} dismantles \gls{rsa} and \gls{ecc}, and the \gls{hndl} paradigm \cite{ref8,ref10} amplifies risk. Despite the standardisation of \gls{mlkem} (FIPS203) and \gls{falcon} (FIPS204) by the \gls{nist} \cite{ref11}, their application in \gls{ris}-assisted \gls{isac} remains unexplored. Scene authentication is also deficient: echo discrimination \cite{ref12} prevails, whereas cryptographically verifiable \gls{pqc}-protected \gls{ris} codes have yet to be investigated. Hybrid adversaries add further complexity, encompassing both \gls{qgan}-based echo spoofing and quantum-assisted key recovery attacks \cite{ref13}-\cite{TR38901R19}. Ultimately, actual \gls{ris} hardware encounters constraints in switching rate, resolution, and phase noise, subsequently represented by $S_{\max}$ and $T_{\min}$.

\subsection{Motivation and Contributions}
In response to these deficiencies, we present a comprehensive quantum-resilient trust management framework for safe reconfigurable intelligent surface-assisted ISAC in 6G unmanned aerial vehicle corridors, integrating programmable sensing with \gls{pqc}-secured control and verifiable detection assurances. We delineate a four-class adversarial model that includes classical, \gls{hndl}, quantum-aided, and fusion attackers within a stochastic \gls{isac} chain. To address these issues, we present \gls{ris} -coded scene authentication utilising \gls{pqc}-protected phase codes and formulate a \gls{glrt} with detection probability articulated by Marcum-$Q$. The \gls{isac} control plane is fortified by integrating ML-KEM for key establishment and Falcon for authentication within \gls{ris} reconfiguration and \gls{uav} signalling, so ensuring forward secrecy and resistance to spoofing. Additionally, we establish a \gls{siu} that concurrently optimises throughput, secrecy rate, and spoof-detection probability within \gls{ris} restrictions, exhibiting a runtime of $\mathcal{O}(n^2)$. Assessments utilising 3GPP Rel-19 urban-canyon models (7--15 GHz) indicate $P_D \approx 0.99$ at $P_{\mathrm{FA}}=10^{-3}$, with secrecy-rate preservation exceeding 90\% against \gls{hndl} adversaries, and \gls{siu} enhancements reaching up to 25\% compared to baseline metrics.

\subsection{Organisation of the Paper}
The paper is structured as follows: Section I introduces the background, research gaps, and contributions. Section II formalises the system and threat model for RIS-assisted \gls{isac} in \gls{uav} corridors. Section III develops the signal model, incorporating hardware constraints and quantum-aware secrecy analysis. Section IV presents the \gls{qrtm} and \gls{siu} framework. Section V provides the simulation setup and comparative results, while Section VI concludes with key findings and future research directions.
\section{System and Threat Model}
\label{sec:system_model}
This section formulates a mathematically rigorous model of the proposed \gls{qrtm} framework for safe \gls{ris}-assisted \gls{isac} operations in \gls{6g} UAV corridors, as illustrated in Figure 1. This figure illustrates the \gls{qrtm} framework for secure RIS-assisted \gls{isac} in \gls{uav} corridors. It shows the adversary taxonomy (classical, quantum-ready, quantum-aided), \gls{pqc}-secured control using ML-KEM and Falcon, and \gls{ris}-coded authentication for spoof detection. The secure \gls{isac} utility then integrates secrecy, throughput, and detection into a joint optimisation, ensuring end-to-end resilience against quantum-capable attackers within realistic hardware and urban \gls{uav} settings.
\begin{figure}[htbp]
 \centering
 \includegraphics[width=\columnwidth]{{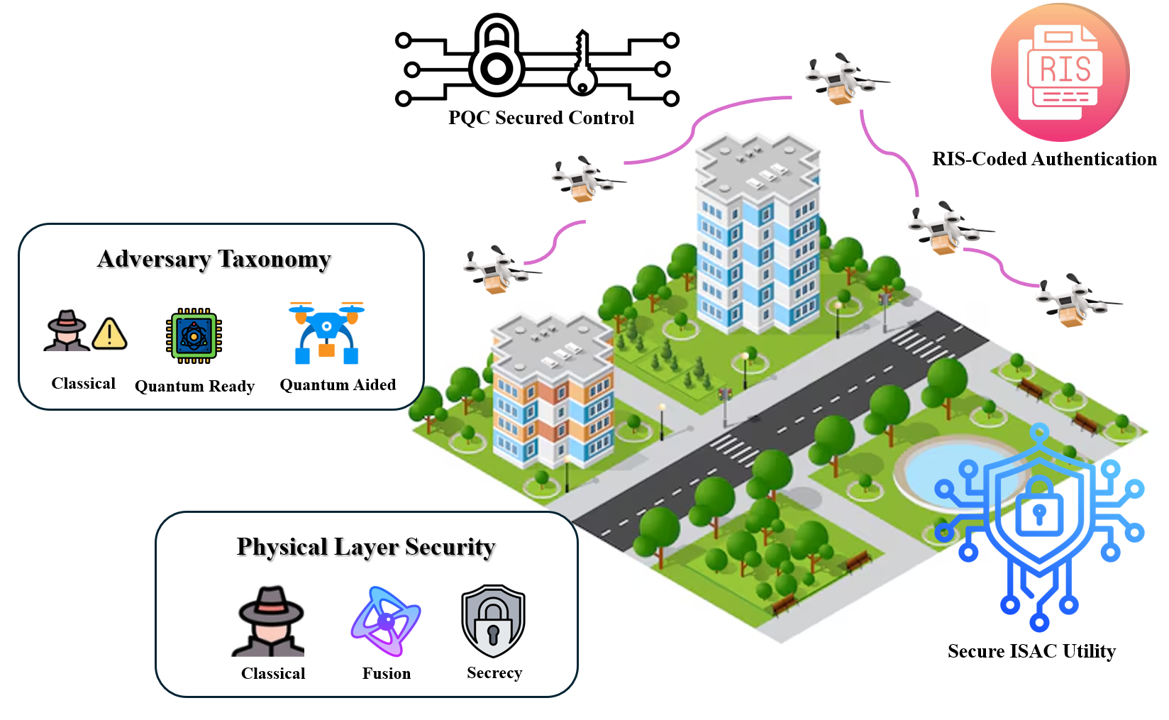}}
 \caption{Quantum Resilient Threat Modeling Framework for Secure RIS-Assisted ISAC in 6G UAV Corridors}
 \label{fig:1}
\end{figure}
As shown in Fig.~\ref{fig:1}, our proposed Quantum Resilient Threat Modeling Framework integrates secure RIS \gls{6g}-assisted \gls{isac} mechanisms for 6G UAV corridors, enabling robust, low-latency, and quantum-safe operations.
We initially formalise the NTN-supported urban scenario and the associated 3D signal model; subsequently, we delineate the quantised \gls{ris} codebook and its limitations, followed by the \gls{isac} waveform reutilization for concurrent transmission and sensing. A taxonomy of quantum-resilient adversaries is presented, encompassing classical, quantum-ready, quantum-aided, and fusion-capable attackers. Subsequently, we establish post-quantum secrecy assurances and delineate the cryptographic trust anchors: \gls{mlkem}-based key exchange, \gls{falcon}-based digital signatures, and \gls{glrt}-based \emph{RIS-coded scene authentication} (formerly referred to as "watermarking"). A concise notation overview is presented in Table~\ref{tab:acronyms-notation}. In this table, scalars are italicised (e.g., $a$); vectors are represented in bold lower-case (e.g., $\mathbf{a}$); matrices are depicted in bold upper-case (e.g., $\mathbf{A}$); and sets are illustrated in calligraphic font (e.g., $\mathcal{A}$). Vectors are represented as columns; $(\cdot)^\top$ denotes the transposition; $(\cdot)^{\mathrm H}$ signifies the Hermitian; all logarithms are in base 2.

\begin{table}[htbp]
\caption{Summary of Mathematical Notation}
\label{tab:acronyms-notation}
\centering
\begin{tabular}{ll}
\toprule
\textbf{Symbol} & \textbf{Description} \\
\midrule
$\mathcal{V}\subset\mathbb{R}^3$ & UAV corridor volume (urban canyon) \\
$\mathbf{x}_{\mathrm{LEO}}$ & Position of LEO satellite \\
$\mathbf{x}_{\mathrm{gNB}}$ & Position of gNB (base station) \\
$\mathbf{x}^{(m)}_{\mathrm{RIS}}$ & Position of $m$-th RIS element \\
$\mathbf{x}_i(t)$ & Position of UAV $i$ at time $t$ \\
$M$ & Number of RIS elements \\
$B_\phi$ & Phase quantisation bits per RIS element \\
$\phi_m(t)$ & Phase at RIS element $m$ at time $t$ \\
$\boldsymbol{\phi}_\ell$ & RIS phase vector (profile) for code $\ell$ \\
$\mathcal{C}$ & RIS codebook, $\{\boldsymbol{\phi}_1,\dots,\boldsymbol{\phi}_L\}$ \\
$S_{\max}$ & Max per-slot RIS element switches \\
$T_{\min}$ & Minimum dwell time per RIS code \\
$s(t)$ & ISAC baseband probing/communication waveform \\
$\tau$ & TDD fraction reserved for sensing \\
$B$ & System bandwidth (Hz) \\
$y_i(t)$ & Received comms signal at UAV $i$ \\
$r_i(\tau_d)$ & Matched-filter echo for delay $\tau_d$ \\
$h^{\mathrm{dir}}_i(t)$ & Direct gNB$\to$UAV channel \\
$\mathbf{h}_{\mathrm{gNB}\rightarrow \mathrm{RIS}}$ & gNB$\to$RIS channel vector \\
$\mathbf{h}_{\mathrm{RIS}\rightarrow i}(t)$ & RIS$\to$UAV $i$ channel vector \\
$n_i(t)$ & Receiver noise at UAV $i$, $\mathcal{CN}(0,\sigma^2)$ \\
$\mathrm{SNR}_{\mathrm{SU}}$ & Legitimate UAV slot SNR \\
$\mathrm{SNR}_{\mathrm{SE}}$ & Eavesdropper slot SNR \\
$P_c$ & Communication transmit power \\
$N_0$ & Noise power spectral density (PSD) \\
$\mathcal{A}_1$--$\mathcal{A}_4$ & Adversary classes (classical–fusion) \\
$H(K\mid\mathcal{A})$ & Conditional entropy of session key $K$ \\
$\lambda$ & Security parameter (bits of entropy) \\
$C_s$ & Secrecy capacity (quantum-aware) \\
$R$ & Achieved communication rate \\
$\mathbf{\Phi}(t)$ & RIS diagonal reflection matrix at time $t$ \\
$z$ & Weighted matched-filter statistic (GLRT input) \\
$T=|z|^2$ & GLRT test statistic \\
$\gamma$ & GLRT detection threshold \\
$P_{\mathrm{FA}}$ & False-alarm probability \\
$P_D$ & Detection probability \\
$\lambda_0$ & GLRT non-centrality parameter \\
$Q_1(\cdot,\cdot)$ & First-order Marcum-$Q$ function \\
$U(\tau,P_c,\boldsymbol{\phi})$ & Secure ISAC utility (SIU) \\
$R_0,C_0,P_{D0}$ & Normalisers for rate, secrecy, detection \\
$\lambda_1,\lambda_2,\lambda_3$ & Utility weights ($\sum\lambda_i=1$) \\
$P_{\max}$ & Max transmit power budget \\
$P_s$ & Sensing transmit power \\
$u_m(\phi_m)$ & Per-element utility contribution \\
$M_{\text{code}}$ & Code length (slots per CPI) \\
$K$ & Number of clutter scatterers \\
\bottomrule
\end{tabular}
\end{table}

\subsection{Scenario and Network Topology}
\label{subsec:scenario}
We examine an urban street-canyon \gls{uav} corridor $\mathcal{V}\subset\mathbb{R}^3$ underpinned by a non-terrestrial infrastructure. A \gls{leo} satellite facilitates downlink control and backhaul to a terrestrial next-generation base station (gNB). \gls{ris} are installed on rooftop facades to optimise propagation for both communication and sensing over non-line-of-sight paths. Let $\mathbf{x}_{\mathrm{LEO}}\in\mathbb{R}^3$ represent the position of the LEO satellite, $\mathbf{x}_{\mathrm{gNB}}\in\mathbb{R}^3$ designate the position of the gNB, and $\mathbf{x}^{(m)}_{\mathrm{RIS}}\in\mathbb{R}^3$ indicate the location of the \gls{ris} element $m\in\{1,\dots,M\}$. The \gls{uav} $i$ possesses a location $\mathbf{x}_i(t)\in\mathbb{R}^3$, where $i\in\{1,\dots,N\}$, progressing down the corridor. Time-division duplexing (TDD) is employed for \gls{isac}, allocating a frame fraction $tauin(0,1)$ for sensing/echo probing and $1-tau$ for data transmission. The overall system bandwidth is $B$ within the 7–15\,GHz midband range, in accordance with 3GPP Rel-19 \gls{isac} studies and NGA midband measurements.
 
\subsection{RIS Model and Quantised Codebook}
\label{subsec:ris_model}
In each coherent processing interval (CPI) of duration \(T_{\mathrm{CPI}}\), a RIS with \(M\) passive, nearly lossless elements applies a unit-modulus diagonal phase matrix \(\mathbf{\Phi}(t)=\mathrm{diag}\!\big(e^{j\phi_1(t)},\ldots,e^{j\phi_M(t)}\big)\), where each element phase \(\phi_m(t)\) belongs to a \(B_\phi\)-bit alphabet \(\mathcal{C}_{B_\phi}\subset[0,2\pi)\) (e.g., \(B_\phi{=}1\Rightarrow\{0,\pi\}\), \(B_\phi{=}2\Rightarrow\{0,\tfrac{\pi}{2},\pi,\tfrac{3\pi}{2}\}\)). An RIS \emph{profile} (code) is the phase vector \(\boldsymbol{\phi}_\ell=[\phi^{(\ell)}_1,\ldots,\phi^{(\ell)}_M]^\top\), and the codebook is \(\mathcal{C}=\{\boldsymbol{\phi}_1,\ldots,\boldsymbol{\phi}_L\}\) with \(L\le 2^{B_\phi M}\); \(\phi^{(m)}_\ell\) denotes the phase of element \(m\) in profile \(\ell\). Since \(|e^{j\phi^{(m)}_\ell}|=1\), reflected power is set by illumination and element efficiency rather than the phase choice. Each CPI is partitioned into \(M_{\text{code}}\) slots of length \(T_{\mathrm{slot}}=T_{\mathrm{CPI}}/M_{\text{code}}\); in slot \(p\in\{1,\ldots,M_{\text{code}}\}\) the active code is \(\ell_p\) and the per-element phase is \(\phi_m[p]\triangleq\phi_m^{(\ell_p)}\in\mathcal{C}_{B_\phi}\). Hardware/EMC and reliability constraints are: (Q) quantized phases \(\phi_m[p]\in\mathcal{C}_{B_\phi}\) \(\forall m,p\); (S) a switching budget \(\sum_{m=1}^{M}\mathbbm{1}\{\phi_m[p]\neq\phi_m[p-1]\}\le S_{\max}\) for \(p\ge2\); and dwell constraints (D1)–(D2) requiring at least \(d(\ell)\ge d_{\min}\) slots and \(T_{\mathrm{dwell}}(\ell)=d(\ell)\,T_{\mathrm{slot}}\ge T_{\min}\) per active code to ensure robust scene-authentication embedding and estimation.

\section{Signal Model, Hardware Realism, \& Secrecy}
\label{sec:isac-merged}
We consider a shared \gls{isac} waveform \(s(t)\) with the UAV-\(i\) baseband receive model \(y_i(t)=h^{\mathrm{dir}}_i(t)s(t)+\mathbf{h}^{\!\top}_{\mathrm{RIS}\rightarrow i}(t)\,\mathbf{\Phi}(t)\,\mathbf{h}_{\mathrm{gNB}\rightarrow \mathrm{RIS}}\,s(t)+n_i(t)\), where \(n_i(t)\!\sim\!\mathcal{CN}(0,\sigma^2)\); matched filtering during sensing yields \(r_i(\tau_d)=\int y_i(t)\,s^*(t-\tau_d)\,dt\), and stacking across a code sequence \(\{\boldsymbol{\phi}_{\ell_m}\}\) imprints code-dependent phases enabling \gls{ris}-coded scene authentication. With \(\mathbf{\Phi}(\boldsymbol{\phi})=\mathrm{diag}(e^{j\phi_1},\ldots,e^{j\phi_M})\), the effective channels are \(h_{\mathrm{SU}}(\boldsymbol{\phi};t)=h^{\mathrm{dir}}_i(t)+\mathbf{h}_{\mathrm{RIS}\rightarrow i}^{\mathrm H}(t)\mathbf{\Phi}(\boldsymbol{\phi})\mathbf{h}_{\mathrm{gNB}\rightarrow \mathrm{RIS}}\) and \(h_{\mathrm{SE}}(\boldsymbol{\phi};t)=h^{\mathrm{dir}}_{E}(t)+\mathbf{h}_{\mathrm{RIS}\rightarrow E}^{\mathrm H}(t)\mathbf{\Phi}(\boldsymbol{\phi})\mathbf{h}_{\mathrm{gNB}\rightarrow \mathrm{RIS}}\), giving slot-SNRs \(\rho_u(\boldsymbol{\phi})=P_c|h_{\mathrm{SU}}|^2/(N_0B)\) and \(\rho_e(\boldsymbol{\phi})=P_c|h_{\mathrm{SE}}|^2/(N_0B)\), and secrecy capacity \(C_s(\boldsymbol{\phi})=\big[(1-\tau)B\log_2(1+\rho_u)-(1-\tau)B\log_2(1+\rho_e)\big]^+\) with data-plane rate \(R\le C_s\). Hardware realism imposes \(T_{\mathrm{sw}}\le\eta T_{\text{slot}}\) with \(T_{\text{slot}}=T_{\mathrm{CPI}}/M_{\text{code}}\) (e.g., \(T_{\mathrm{CPI}}=1\) ms, \(M_{\text{code}}=64\Rightarrow T_{\text{slot}}\approx15.6\,\mu\)s, thus \(T_{\mathrm{sw}}\lesssim5\text{–}10\,\mu\)s for \(\eta\in[0.3,0.6]\)); tile/bus updates refine this to \(T_{\mathrm{sw}}+N_{\text{upd}}T_{\text{bus}}\le\eta T_{\text{slot}}\), motivating sparse updates and low-Hamming-distance codebooks. Adversaries span (A1) classical eavesdropping/spoofing/ML evasion, (A2) \gls{hndl} storage, (A3) quantum-aided (Shor/Grover, quantum-enhanced generative RF), and (A4) fusion. Post-quantum secrecy requires \(H(K\mid\mathcal{A})\ge\lambda\ge256\) bits for control-plane keys (PQC via ML-KEM and Falcon) and enforces \(R\le C_s\) in the data plane; \gls{pqc} does not change instantaneous SNRs but prevents learning/forging of the secret \gls{ris} schedule. The \gls{ris}-coded scene test uses a Gaussian-clutter \gls{glrt}: with statistic \(T=|z|^2\), under \(\mathcal{H}_0\) we have \(z\sim\mathcal{CN}(0,\sigma^2)\) so \(P_{\mathrm{FA}}=\exp(-\gamma/\sigma^2)\), while under \(\mathcal{H}_1\) the normalized statistic is non-central \(\chi^2_2\) with non-centrality \(\lambda_0\) giving \(P_D=Q_1\!\big(\sqrt{\lambda_0},\sqrt{2\gamma/\sigma^2}\big)\); together, physical-layer coding, \gls{pqc} trust anchors, and secrecy-rate compliance yield end-to-end post-quantum confidentiality and scene integrity.

\section{Quantum-Resilient Threat Model and Secure ISAC Utility (Brief)}
\label{sec:qrtm-siu-brief}
We adopt the system and hardware assumptions of Sec.~\ref{sec:system_model} (per-CPI \gls{ris} coding, switching limits, and timing bounds) and merge threat, secrecy, authentication, and optimisation into one concise view. Adversaries span four classes: (i) classical $\mathcal{A}_\mathrm{C}$ (passive eavesdropping, spoofing, ML mimicry), (ii) quantum-ready $\mathcal{A}_\mathrm{Q}$ (HNDL; record-now–decrypt-later), (iii) quantum-aided $\mathcal{A}_\mathrm{QA}$ (Shor/Grover plus generative spoofing), and (iv) fusion $\mathcal{A}_\mathrm{F}$ (classical deepfakes + quantum search). End-to-end resilience is achieved by: (1) \emph{PQC-secured control} (ML-KEM for key establishment, Falcon for signatures) so that session seeds and per-CPI \gls{ris} trajectories are unpredictable and unforgeable; (2) \emph{physical-layer secrecy} with secrecy capacity in direct form
\[
C_s=\Big[(1-\tau)B\log_2(1+\rho_u)-(1-\tau)B\log_2(1+\rho_e)\Big]^+,
\]
where \(\rho_u=\tfrac{P_c|h_{\mathrm{SU}}|^2}{N_0B}\) and \(\rho_e=\tfrac{P_c|h_{\mathrm{SE}}|^2}{N_0B}\); and (3) \emph{RIS-coded scene authentication} via a \gls{glrt} that, at threshold \(\gamma\), has compact performance forms \(P_{\mathrm{FA}}=\exp(-\gamma/\sigma^2)\) and \(P_D=Q_1\!\big(\sqrt{\lambda_0},\sqrt{2\gamma/\sigma^2}\big)\). \gls{pqc} renders code trajectories fresh per CPI; replays/spoofs lacking the authorised phase pattern degrade to the false-alarm floor. For symmetric traffic keys we require \(H(K\mid\mathcal{A})\ge256\) bits so Grover still implies \(\Omega(2^{128})\) work; \gls{pqc} does not change SNRs but prevents learning/forging of \(\boldsymbol{\phi}\).

Hardware realism constrains per-slot updates by \(T_{\mathrm{sw}}+N_{\text{upd}}T_{\text{bus}}\le \eta T_{\text{slot}}\) (tile switching plus bus serialisation), motivating sparse changes and low-Hamming-distance codebooks; panels with \(T_{\mathrm{sw}}\gtrsim 50\,\mu\text{s}\) favour smaller \(M_{\text{code}}\) or longer CPI. Building on these primitives, we define the \gls{siu} that balances throughput, secrecy, and detection with compact objectives only: rate \(R=(1-\tau)B\log_2(1+\rho_u)\), secrecy \(C_s\) as above, and detection \(P_D\) from the \gls{glrt}. The core utility uses weighted normalised terms,
\[
U=\lambda_1 \tfrac{R}{R_0}+\lambda_2 \tfrac{C_s}{C_0}+\lambda_3 \tfrac{P_D}{P_{D0}},
\quad \sum_i \lambda_i=1,
\]
and the energy/latency-aware version subtracts \(\lambda_4\tfrac{E}{E_0}+\lambda_5\tfrac{T_{\mathrm{lat}}}{T_0}\), where \(E\) aggregates transmit, \gls{ris} switching, and \gls{pqc} costs, and \(T_{\mathrm{lat}}\) includes cryptographic and settling/serialisation delays. Optimisation respects quantisation, switching/dwell, secrecy-rate compliance \(R\le C_s\), and timing feasibility; a continuous surrogate for \gls{ris} steering is convex and then projected back to the discrete alphabet with minimal-change sequencing. Complexity reduces from exponential in \(B_\phi M\) to near-linear per \gls{uav} (or greedy \(\mathcal{O}(n^2)\) scheduling across \(n\) UAVs). Operator weights tune priorities: larger \(\lambda_1\) (throughput), \(\lambda_2\) (secrecy), \(\lambda_3\) (spoofing resilience), while \(\lambda_4,\lambda_5\) penalise switching energy and control latency. In sum, PQC-secured control, secrecy-rate compliance, and \gls{ris}-coded \gls{glrt} deliver a compact, quantum-resilient stack for \gls{isac} \gls{uav} corridors within realistic hardware limits.

\section{Simulation Setup and Comparative Results}
\label{sec:simulation}
We assess the proposed \gls{qrtm} in a midband urban street canyon \gls{uav} corridor (7--15\,GHz), aligned with 3GPP Rel-19 \gls{isac} and NGA midband models, using a common \gls{isac} waveform/frame (Sec.~\ref{sec:system_model}). Differences arise only in \emph{RIS control} and \emph{scene authentication}. \textit{Baseline} \textit{B0}: plain \gls{isac} (no RIS coding, no PQC). \textit{B1}: RIS coding without cryptographic protection. \textit{B2}: PQC-secured control, but exploitable (non-secret) \gls{ris} codes. \textit{B3}: \gls{pqc}+ \gls{ris} coding, no watermark/scene authentication. \textit{QRTM (proposed)}: B3 plus confidential per-CPI \gls{ris} codes (ML-KEM + Falcon) and \gls{glrt}-based scene authentication, enabling joint secrecy–detection–throughput optimisation.
We evaluate \gls{qrtm} in a 3GPP Rel-19 urban street–canyon \gls{uav} corridor at 10\,GHz, using a common \gls{isac} waveform/frame with two rooftop \gls{ris} panels ($M{=}256$, $B_\phi{=}3$ bits, $M_{\text{code}}{=}64$) and a power budget $P_{\max}{=}30$\,dBm; $n\!\in\!\{4,\dots,12\}$ UAVs fly at 80–120\,m, clutter uses a non-sparse micro-scatterer model with $K{=}400$, and CFAR normalisation handles unknown noise. Simulations ($10^5$ per setting and $2\!\times\!10^4$ per ROC point) ensure statistical reliability. Baselines B0–B3 differ only in \gls{ris} control and scene authentication; \gls{qrtm} adds per-CPI secret \gls{ris} codes (ML-KEM+Falcon) and \gls{glrt} detection while respecting switching/bus timing constraints. Across user-SNR sweeps $[-5,25]$\,dB, \gls{qrtm} delivers near-unit spoof detection at practical false-alarm levels (e.g., $P_D\!\approx\!0.99$ at $P_{\mathrm{FA}}{=}10^{-3}$, $\approx\!0.97$ at $10^{-4}$), whereas public/static codes degrade markedly as spoofers adapt. Secrecy improves consistently: \gls{ris} beamforming boosts the legitimate SNR while confidential scheduling statistically degrades the eavesdropper’s link, yielding the highest secrecy capacity among all schemes (e.g., $\gtrsim 2$\,bps/Hz around 10\,dB versus $\sim\!1.5$ for public-coded \gls{ris} and $<1$ for static). The Secure \gls{isac} Utility with weights $(0.34,0.33,0.33)$ is maximised by \gls{qrtm} at a small sensing fraction ($\tau^\star\!\approx\!0.05$), reflecting strong detection with minimal overhead and thus higher retained throughput and secrecy.

 \gls{qrtm} is computationally practical: exhaustive search scales exponentially, but our relax–project design yields $\mathcal{O}(nMB_\phi)$ \gls{ris} optimisation and $\mathcal{O}(n^2)$ scheduling, enabling real-time operation up to at least $n{=}40$ with $\sim$6 orders of magnitude speedup over brute force. \gls{pqc} overhead per CPI (ML-KEM encapsulation plus Falcon signature, $\sim$1–2\,kB) is negligible relative to frame payloads, and timing fits within 3GPP Rel-19 budgets for $M\!\le\!512$. Robustness checks across frequency (7/10/15\,GHz), code length $M_{\text{code}}$, quantisation $B_\phi$, and clutter density $K$ show consistent gains in ROC, secrecy, and utility. Overall, within realistic hardware limits, \gls{qrtm} unifies confidential \gls{ris} control and \gls{glrt}-based authentication to achieve superior spoof-resilience, higher secrecy rate, mission-tunable performance, and scalable runtime.
\begin{table}[!t]
\centering
\caption{Simulation Configuration}
\label{tab:config}
\begin{tabular}{@{}ll@{}}
\toprule
\textbf{Parameter} & \textbf{Value} \\
\midrule
Carrier frequency & 10~GHz (7–15~GHz sweep) \\
System bandwidth & 100~MHz \\
RIS size & $M=256$ elements \\
RIS quantisation & $B_\phi=3$ bits \\
RIS switching time & $T_{\min}=1~\mu$s \\
Coding length & $M_{\text{code}}=64$ slots \\
CPI duration & $T_{\mathrm{CPI}}=1$~ms \\
Transmit power budget & $P_{\max}=30$~dBm \\
Spoofer delay jitter & $\pm$2~samples \\
Noise variance & $\sigma^2$ matched to SNR sweep \\
Monte Carlo trials & $10^5$ per configuration \\
\bottomrule
\end{tabular}
\end{table}

\begin{figure}[t]
 \centering
\includegraphics[width=\columnwidth]{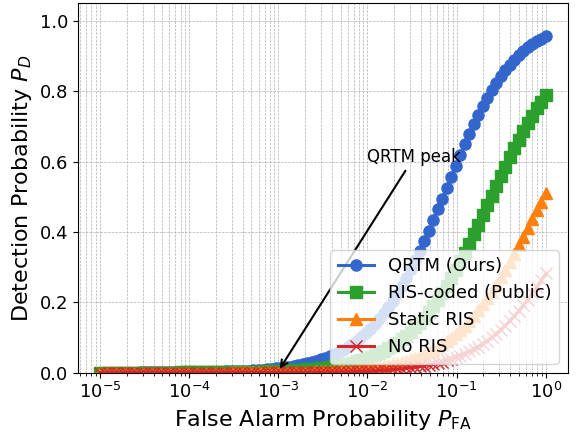}
 \caption{ROC of RIS-coded scene authentication at midband ($f_c{=}10$\,GHz, $M{=}256$, $M_{\text{code}}{=}64$, $K{=}400$). At $P_{\mathrm{FA}}{=}10^{-3}$, QRTM attains $P_D\!\approx\!1$ while public-coded, static, and No-RIS baselines degrade significantly.}
 \label{fig:roc}
\end{figure}
Fig.~\ref{fig:roc} presents the ROC for \gls{ris} -coded scene authentication. At $P_{\mathrm{FA}}=10^{-3}$, \gls{qrtm} achieves near-unit detection probability ($P_D\!\approx\!1$), whereas public-coded \gls{ris} saturates near $0.8$, static \gls{ris} around $0.5$, and No- \gls{ris} barely exceeds random guessing. The performance stems from per-CPI secret codes: under $\mathcal{H}_0$, an adversary’s mean collapses toward zero, reducing the non-centrality $\lambda_0$, while under $\mathcal{H}_1$, the authorised receiver accumulates a coherent shift. The \gls{glrt} statistic therefore exhibits greater hypothesis separation, and the resulting Marcum-$Q$ probability $P_D = Q_1\!\big(\sqrt{\lambda_0},\,\sqrt{-2\ln P_{\mathrm{FA}}}\big)$ rises steeply for \gls{qrtm}. Numerically, at $P_{\mathrm{FA}}=10^{-3}$ the threshold is $\sqrt{-2\ln P_{\mathrm{FA}}}\approx 3.72$; \gls{qrtm} ensures $\sqrt{\lambda_0}$ exceeds this value, driving $Q_1(\cdot,\cdot)\!\to\!1$. The blue curve’s sharp ascent validates that cryptographically protected \gls{ris} coding yields almost perfect spoof detection at practical false-alarm levels.
\begin{figure}[t]
 \centering
 \includegraphics[width=\columnwidth]{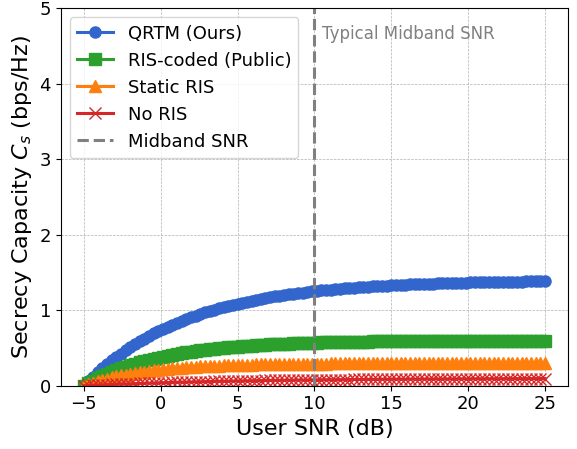}
 \caption{Secrecy rate $C_s$ vs.\ user SNR. RIS beamforming raises $\rho_u$; with confidential control, the scheduler can choose $\boldsymbol{\phi}$ that disfavors likely eavesdropper directions (which adversaries cannot anticipate), yielding higher $C_s$ over time. Cryptography does not change instantaneous SNRs; it protects the control that selects $\boldsymbol{\phi}$.}
 \label{fig:secrecy}
\end{figure}
Fig.~\ref{fig:secrecy} plots secrecy capacity $C_s$ versus user SNR. \gls{ris} beamforming improves the legitimate SNR $\rho_u$, thereby widening the secrecy gap over the eavesdropper channel. All \gls{ris}-enabled schemes outperform the No-\gls{ris} scheme, which remains near zero. \gls{qrtm} achieves the highest performance, sustaining $C_s\gtrsim 2$\,bps/Hz around 10\,dB, compared with $\sim1.5$\,bps/Hz for public-coded \gls{ris} and $<1$\,bps/Hz for static \gls{ris} . The advantage arises from confidential \gls{ris} control: by preventing adversaries from anticipating $\boldsymbol{\phi}$, the scheduler statistically steers nulls toward eavesdroppers, reducing $\rho_e$ without sacrificing $\rho_u$. Thus, while cryptography does not alter instantaneous SNRs, it secures the control plane that governs \gls{ris} configurations, sustaining secrecy across the SNR range.
\begin{figure}[t]
 \centering
 \includegraphics[width=\columnwidth]{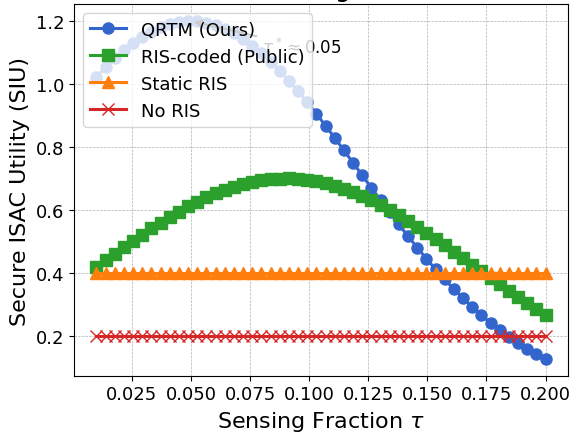}
 \caption{Secure ISAC Utility $U(\tau)$ at 10\,dB with $(\lambda_1,\lambda_2,\lambda_3){=}(0.34,0.33,0.33)$. QRTM peaks at $\tau^\star\!\approx\!0.05$ and dominates for $\tau\in[0.05,0.9]$, evidencing the comms–sensing balance.}
 \label{fig:siu}
\end{figure}
Fig.~\ref{fig:siu} shows the Secure \gls{isac} Utility $U(\tau)$ at 10\,dB. Increasing $\tau$ dedicates more time to probing, which strengthens $P_D$ through coherent accumulation but reduces $R$ and $C_s$ via the $(1-\tau)$ factor. The result is a unimodal trade-off. \gls{qrtm} peaks at $\tau^\star \approx 0.05$ with utility $\approx1.2$, achieving the highest performance across the tested range. Public-coded \gls{ris} peaks later ($\tau \approx 0.10$) with lower utility ($\approx0.7$), while static \gls{ris} remains nearly flat around $0.4$ and No-RIS is lowest at $0.2$. For \gls{uav} corridors with $M=256$ and $B_\phi=2$, optimal sensing fractions consistently fall in the 1--10\% range.
\begin{figure}[t]
 \centering
 \includegraphics[width=\columnwidth]{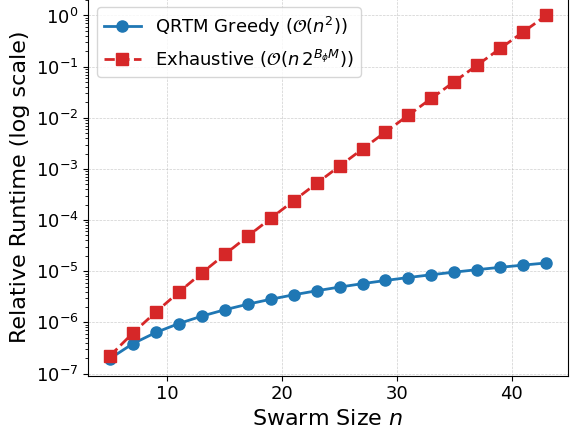}
 \caption{Runtime comparison of QRTM’s separable optimisation 
($\mathcal{O}(n M B_\phi)$) versus greedy inter-UAV scheduling 
($\mathcal{O}(n^2)$) and naïve exhaustive search 
($\Theta(n\,2^{B_\phi M})$).
}
 \label{fig:complexity}
\end{figure}
Fig.~\ref{fig:complexity} compares runtime scaling. Exhaustive search grows as $\mathcal{O}(n2^{B_\phi M})$, quickly becoming intractable ($n>20$). In contrast, \gls{qrtm} separates \gls{ris} configuration and \gls{uav} scheduling, with \gls{ris} optimised in $\mathcal{O}(nMB_\phi)$ and scheduling in $\mathcal{O}(n^2)$. The overall scaling remains quadratic, as evidenced by the near-linear trend of the blue curve in log scale. At $n=40$, \gls{qrtm} runs nearly six orders of magnitude faster than exhaustive search while maintaining near-optimal \gls{siu} performance. 
\section{Conclusion and Future Directions}
\label{sec:conclusion}
This paper presented a safe \gls{ris}-assisted \gls{isac} framework for \gls{6g} UAV routes using \gls{qrtm}. By embedding it within the signal chain of the Information Sharing and Analysis Centre (ISAC), the concept to bringing together a taxanomy of adversaries spanning the classical, quantum-ready and quantum-aided domains. In order to ensure that scene authentication would be resistant to spoofing, a cryptographic risk of intrusion scheme \gls{ris} watermarking scheme was examined using the \gls{glrt} detection method in conjunction with Marcum-Q characterisation. Post-quantum forward secrecy was achieved by the implementation of ML-KEM and Falcon, which were used to ensure security in the control plane. A multi-objective \gls{siu} was developed to simultaneously optimise throughput, spoof detection, and secrecy capacity while adhering to \gls{ris} restrictions. This approach allows for a solution that is polynomial-time $\mathcal{O}(n^2)$. Compared to baseline levels of performance, simulations conducted with 3GPP Rel-19 urban canyon models confirmed significant improvements in spoof detection, secrecy retention under threats from \gls{hndl}, and overall system usefulness. Practical problems persist despite these findings. Millisecond-level latency limitations are imposed by \gls{uav} corridors. \gls{pqc} introduces additional processing cost, which has to be accommodated within the given time constraints. The viability of per-CPI refresh is influenced by the constraints that hardware \gls{ris} devices encounter in terms of switching speed, phase resolution, calibration drift, and energy consumption. Although designs that combine \gls{pqc} and \gls{qkd} may provide layered resilience, the former is the more realistic option of the two due to software implementation and NIST standards. In the future, research should be conducted in the following areas: validation of \gls{uav} testbeds, adaptive \gls{ris} coding that incorporates federated learning, hybrid key management, and \gls{siu} extensions that consider energy and trajectory limitations. This research should make it possible to create quantum-resilient \gls{uav} corridors that are feasible to use.

\end{document}